\documentclass[conference]{IEEEtran}
\usepackage{graphicx}
\usepackage{amsmath, amssymb}
\usepackage{bbm}
\usepackage{cite}
\usepackage{mathtools}
\usepackage{caption}
\usepackage{subcaption}
\usepackage{bbold}
\usepackage{dsfont}
\usepackage{amsfonts}
\usepackage{tikz}
\usepackage{authblk}
\usepackage{enumerate}
\usepackage[caption=false,font=footnotesize]{subfig}

\usetikzlibrary{arrows,automata,positioning}

\usepackage{xcolor}

\title{Relay Selection and User Equipment Admission in Resource-Efficient  NextG Sidelink Communications}
\vspace{-2cm}
\author{Yalin E. Sagduyu, Tugba Erpek, Sastry Kompella, and Kemal Davaslioglu
\\ {\normalsize  Nexcepta, Gaithersburg, MD, USA} \thanks{
This material is based upon work supported by the Office of Under Secretary of Defense, Research and Engineering, under contract (OTA\#) W900KK-23-9-0034.}}

\date{}
\IEEEoverridecommandlockouts
\begin{document}

\maketitle
\vspace{-2cm}
\thispagestyle{empty}

\begin{abstract}
5G/6G sidelink communications addresses the challenge of connecting outer UEs, which are unable to directly access a base station (gNodeB), through inner UEs that act as relays to connect to the gNodeB. The key performance indicators include the achievable rates, the number of outer UEs that can connect to a gNodeB, and the latency experienced by outer UEs in establishing connections. We consider problem of determining the assignment of outer UEs to inner UEs based on the channel, interference, and traffic characteristics. We formulate an optimization problem to maximize a weighted sum rate of UEs, where weights can represent priority, waiting time, and queue length. This optimization accommodates constraints related to channel and interference characteristics that influence the rates at which links can successfully carry assigned traffic. While an exhaustive search can establish an upper bound on achievable rates by this non-convex optimization problem, it becomes impractical for larger number of outer UEs due to scalability issues related to high computational complexity. To address this, we present a greedy algorithm that incrementally selects links to maximize the sum rate, considering already activated links. This algorithm, although effective in achieving high sum rates, may inadvertently overlook some UEs, raising concerns about fairness. To mitigate this, we introduce a fairness-oriented algorithm that adjusts weights based on waiting time or queue length, ensuring that UEs with initially favorable conditions do not unduly disadvantage others over time. We show that this strategy not only improves the average admission ratio of UEs but also ensures a more equitable distribution of service among them, thereby providing a balanced and fair solution to sidelink communications.
\end{abstract}


\section{Introduction}
In the rapidly advancing landscape of mobile communications, the emergence of \emph{5G networks} and the anticipated transition to \emph{6G} represent significant technological advances. These networks are set to support an Unparalleled range of applications by enhancing broadband connectivity, enabling a massive influx of connected devices through the Internet of Things (IoT), and empowering new technologies across industries such as defense, public safety, and automotive. Despite these innovations, ensuring \emph{consistent and reliable network coverage} remains a significant challenge, particularly for user equipments (UEs) located at the network's edge or in underserved areas with poor connectivity to a base station (gNodeB).

\begin{figure}[ht!]
	\centering
	\includegraphics[width=\columnwidth]{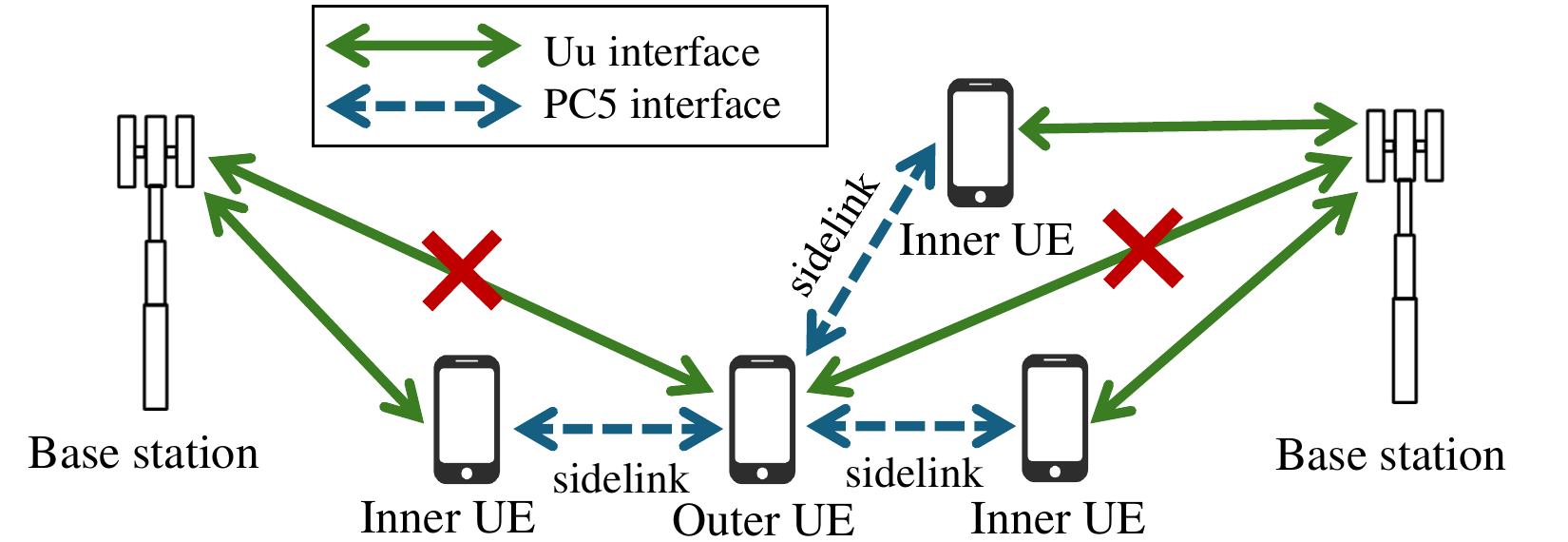}
	\caption{Sidelink communications.}
	\label{fig:system}
\end{figure}

\emph{Sidelink communications} plays a pivotal role in addressing these challenges of  5G/6G by enabling direct \emph{device-to-device (D2D)} interactions \cite{weerackody2023needs, harounabadi2021v2x, lien20203gpp, ganesan20235g}. 
As shown in Fig.~\ref{fig:system}, sidelink allows UEs to communicate with each other (through PC5 interface) and with gNodeB (through Uu interface) without intermediary infrastructure such as in vehicle-to-everything (V2X) networks, enhancing the network's reach and reliability. The significance of sidelink communications becomes particularly evident in situations where outer UEs cannot directly connect to the base station due to their geographical locations or the physical limitations of the network. The architectural design of \emph{multi-hop} sidelink communications addresses the limitations inherent in the range and reach of traditional single-hop network systems. By strategically leveraging inner UEs that have robust connections to the gNodeB, the network can extend its service coverage to include outer UEs 
\cite{shrivastava2023sidelink, bazzi2021design, elkadi2023open, gamboa2023system, wang2024sideseeker,fu2023generalized, decarli2024performance,weerackody2024nr}.

Sidelink communications is highly important for next-generation communication systems for several reasons: (i) \emph{Enhanced connectivity}: Sidelink allows devices to communicate directly without relying on a centralized network infrastructure, which is crucial in environments where traditional communication networks are unavailable or unreliable. (ii) \emph{Resilience and reliability}: Sidelink can maintain communication links even when parts of the network are compromised or under (interception or disruption) attack. (iii) \emph{Reduced latency}: D2D communication can help reduce delays in transmitting crucial operational data, enhancing the responsiveness of communication devices. (iv) \emph{Network scalability and flexibility}: Sidelink communications can dynamically adapt to changing conditions and requirements, which is essential for operations that involve rapid deployment and mobility. 

To facilitate efficient sidelink communications, an effective mechanism for \emph{associating UEs over multiple hops} is crucial. In practical scenarios, the association of outer UEs to inner UEs, which act as relays, ensures that devices beyond the range of a gNodeB can still access the network. This multi-hop connectivity not only extends the network's coverage but also optimizes the use of spectral resources, thereby improving the overall network performance in terms of achievable rates by the UEs, latency (time taken for UEs to connect to the gNodeB potentially via relays), and the admission ratio of UEs connecting to the gNodeB.

We formulate the selection of UE-UE and UE-gNodeB associations as an \emph{optimization problem}. The goal is to maximize the weighted sum rate, where weights may reflect various factors such as priority, waiting time, or queue length. The problem is constrained by the traffic capabilities and channel conditions, including interference considerations, that influence the achievable rates of UEs over multiple hops.

To address the computational complexity inherent in solving this non-convex optimization problem, particularly for large sets of UEs, we propose two main algorithmic solutions: (i) a \emph{greedy algorithm} that incrementally selects the most beneficial links to maximize the sum rate, and (ii) a \emph{fairness-oriented algorithm} that dynamically adjusts the weights of UEs to ensure equitable access to network resources. Greedy algorithm focuses on incrementally building an effective network by selecting and activating the most beneficial links at each step, thereby maximizing the immediate gain in weighted sum rate. However, this may lead to the \emph{fairness} issue regarding the diversity of UEs served. Recognizing the potential for service inequities in a greedy approach, fair algorithm adjusts priorities dynamically based on the accrued \emph{waiting time} or \emph{queue length} of UEs. This ensures a more balanced and equitable distribution of network resources among all UEs. Note that the globally optimal solution is not computationally feasible as the network size grows. A \emph{distributed} approach can be followed for relay association in sidelink communications. We show that the proposed greedy algorithm sustains high levels of sum rate but may not activate all UEs, leading to fairness issues. On the other hand, the proposed fair algorithm reduces the variance among admission ratios of different UEs while sustaining high levels of sum rate and mean admission ratio for UEs supported in sidelink communications.   

The remainder of the paper is organized as follows. Sec.~\ref{sec:system_model} describes the system model and presents the optimization problem. Sec.~\ref{sec:global} analyzes the global optimization performance and points at computational limitations. Secs.~\ref{sec:greedy}, \ref{sec:fair}, and \ref{sec:distributed} present greedy, fairness-oriented, and distributed solutions, respectively. Sec.~\ref{sec:conclusion} concludes the paper.

\section{System Model} \label{sec:system_model}

\begin{figure}[ht!]
	\centering
	\includegraphics[width=0.6\columnwidth]{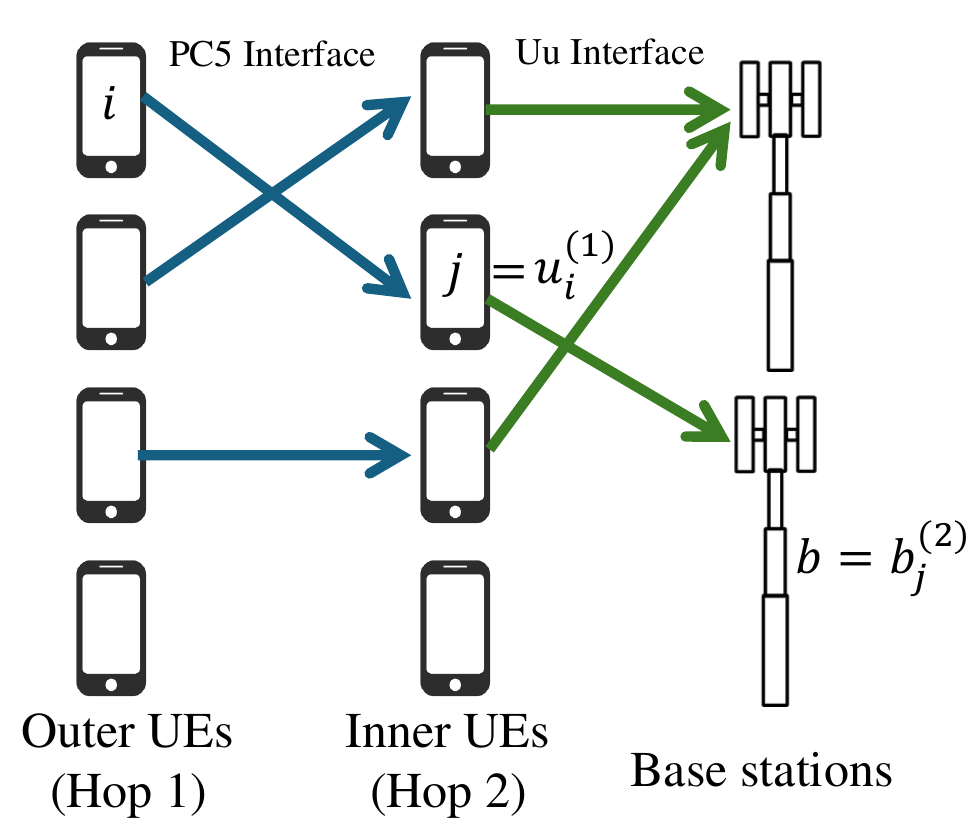}
	\caption{Relay selection for sidelink communications.}
	\label{fig:system2}
\end{figure}

The goal is to select the relay UE(s) based on the channel and traffic characteristics, as shown in Fig.~\ref{fig:system2}. We assume that Hop 2 (inner or relay) UEs can reach a gNodeB but Hop 1 (outer) UEs cannot (they can only reach Hop 2 UEs). Each UE $j$ in Hop 2 is already assigned to gNodeB $b_j^{(2)}$. The problem is to determine the assignment of UE $i$ in Hop 1, $\mathcal{H}_1$, to UE $u_i^{(1)}$ in Hop 2, $\mathcal{H}_2$. We have the following relay assignment constraints: Any UE in Hop 2 with single interface can only be connected to one UE $i$ in $\mathcal{H}_1$ such that $u_i^{(1)} \neq u_{i'}^{(1)}$ for any $i \neq i'$. It is possible that a UE $i$ in $\mathcal{H}_1$ is not connected to any UE in Hop 2 such that $u_i^{(1)} = \emptyset$. UEs in $\mathcal{H}_1$ that are simultaneously activated would interfere with each other. A gNodeB (with multiple antennas) could serve multiple UEs in Hop 2 without interference. We have the following link rate and traffic rate definitions. $c_{ij}^{(1)}$ is the link rate (capacity) from UE $i$ in $\mathcal{H}_1$ to UE $j \in \mathcal{H}_2$. $c_{jb}^{(2)}$ is the link rate (capacity) from UE $j \in \mathcal{H}_2$ to gNodeB $b$. $r_i^{(1)}$ is the traffic rate of UE $i$ in $\mathcal{H}_1$. $r_j^{(2)}$ is the traffic rate of UE $j \in \mathcal{H}_2$. 

The two bounds on traffic rates are given as follows.
\begin{enumerate}[{Hop 1:}]
	\item $r_i^{(1)} \leq c_{i,j}^{(1)}$ for $i \in \mathcal{H}_1$ and any UE $j=u_i^{(1)} \in \mathcal{H}_2$,
	\item  $r_i^{(1)} + r_j^{(2)} \leq c_{j,b_j^{(2)}}^{(2)}$ for any UE $i$ in $\mathcal{H}_1$ and any UE $j = u_i^{(1)}$ in $\mathcal{H}_2$.
\end{enumerate}

Link rate can be expressed as a function of signal-to-interference-and-noise ratio (SINR) for different hops:
\begin{enumerate} [{Hop 1:}]
	\item  $c_{ij}^{(1)} = \log \left(1+ \frac{1(u_i^{(1)} \neq \emptyset) P_i  h_{ij}^{(1)}}{\sigma^2  + \sum_{k \in \mathcal{H}_1, k \neq i} 1(u_k^{(1)} \neq \emptyset) P_k  h_{kj}^{(1)} } \right) $,
	\item $c_{jb}^{(2)} = \log \left( 1+ \frac{P_j  h_{j b}^{(2)} } {\sigma^2} \right)$,
\end{enumerate} 
where $h_{ij}^{(1)}$ is channel from UE $i \in \mathcal{H}_1$ to UE $j \in \mathcal{H}_2$, $h_{jb}^{(2)}$ is channel from UE $j \in \mathcal{H}_2$ to gNodeB $b$, and $P_i$ and $P_j$ are the transmit powers of Outer UE $i$ and Inner UE $j$, respectively. We assume PC5 and Uu interfaces operate at different frequency bands so that Hop 1 and Hop 2 UEs do not interfere. Given $h_{ij}^{(1)}$, $h_{i,b_j^{(2)}}^{(2)}$, $r_j^{(2)}$ and $b_j^{(2)}$ for any $i \in \mathcal{H}_1$ and $j \in \mathcal{H}_2$, the optimization problem is formulated as selecting $u_i^{(1)}$  (the assignment of any UE $i \in \mathcal{H}_1$ to a UE in $\mathcal{H}_2$) as follows:
\\

\noindent \textbf{maximize:} $\sum_{i \in \mathcal{H}_1 } w_i r_i^{(1)}$

\noindent \textbf{variables:} $\{u_i^{(1)}, i \in \mathcal{H}_1 \}$ 

\noindent  \textbf{constraints:}
\begin{enumerate}[(C1)]
\item $r_i^{(1)} \leq c_{i,j}^{(1)}$ and any UE $j=u_i^{(1)} \in \mathcal{H}_2$, \\ \item $r_i^{(1)} + r_j^{(2)} \leq c_{j,b_j^{(2)}}^{(2)}$ for any UE $i \in \mathcal{H}_1$ and any UE $j = u_i^{(1)} \in \mathcal{H}_2$, \\
\item $u_i^{(1)} \neq u_{i'}^{(1)}$ for any $i \neq i'$, \\
\item $c_{ij}^{(1)} = \log \left(1+ \frac{1(u_i^{(1)} \neq \emptyset) P_i  h_{ij}^{(1)}}{\sigma^2  + \sum_{k \mathcal{H}_1, k \neq i} 1(u_k^{(1)} \neq \emptyset) P_k  h_{kj}^{(1)} } \right)$,
\item $c_{jb}^{(2)} = \log \left( 1+ \frac{P_j  h_{j b}^{(2)} } {\sigma^2} \right)$.
\end{enumerate}

We will evaluate the solution to this optimization problem in the following sections, first as a globally optimal solution in Sec.~\ref{sec:global} and then with low-complexity greedy and fair algorithms in Secs.~\ref{sec:greedy} and \ref{sec:fair}, respectively, followed by the discussion of distributed operation in Sec.~\ref{sec:distributed}.

\section{Global Solution \label{sec:global}}

\begin{figure}[ht!]
	\centering
	\includegraphics[width=0.5\columnwidth]{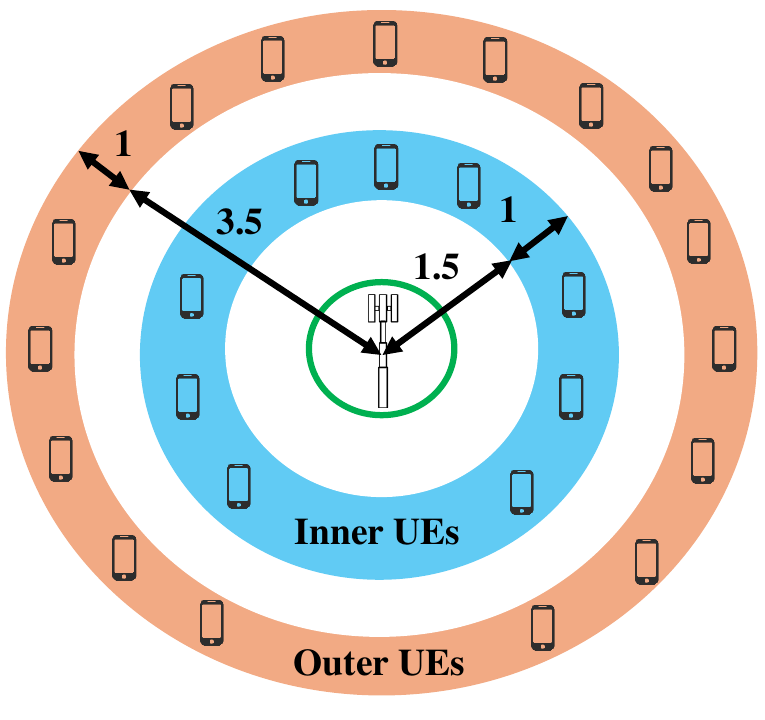}
	\caption{Network topology.}
	\label{fig:topology}
\end{figure}

The optimization problem formulated in Sec.~\ref{sec:system_model} is not convex and involves integer programming. To obtain initial results (that may serve as an upper limit), we perform an exhaustive search over possible link schedules. The base station is positioned at the center, with inner nodes randomly distributed within an annular region bounded by an inner radius of 1.5 and an outer radius of 2.5, while outer nodes are randomly distributed within a second annular region defined by an inner radius of 3.5 and an outer radius of 4.5 as shown in Fig.~\ref{fig:topology}. We vary the number of outer UEs denoted by $n_o$, while we assume the same number for outer and inner UEs for numerical results. To start with, we set path loss coefficient to 2, weights of UEs to 1, and assume that there is no relay traffic. For a given $n_o$, we generate 1000 channel conditions. For each channel condition, we search for the best achievable rate over up to 50000 link schedules and average the best rates over all channel conditions. 

\begin{figure}[t!]
	\centering
	\vspace{-0.4cm}
	\includegraphics[width=0.8\columnwidth]{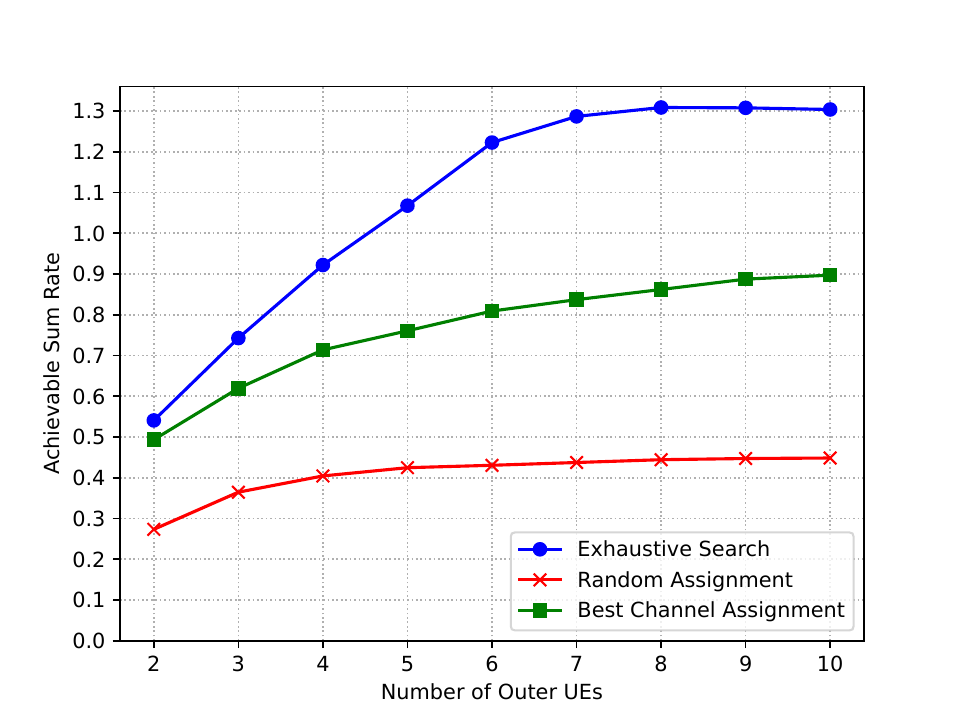}
	\caption{Sum rate achieved by exhaustive search, random assignment, and best channel assignment.}
	\label{fig:sumrate_exhaustive_Random_bc}
		\centering
		\includegraphics[width=0.8\columnwidth]{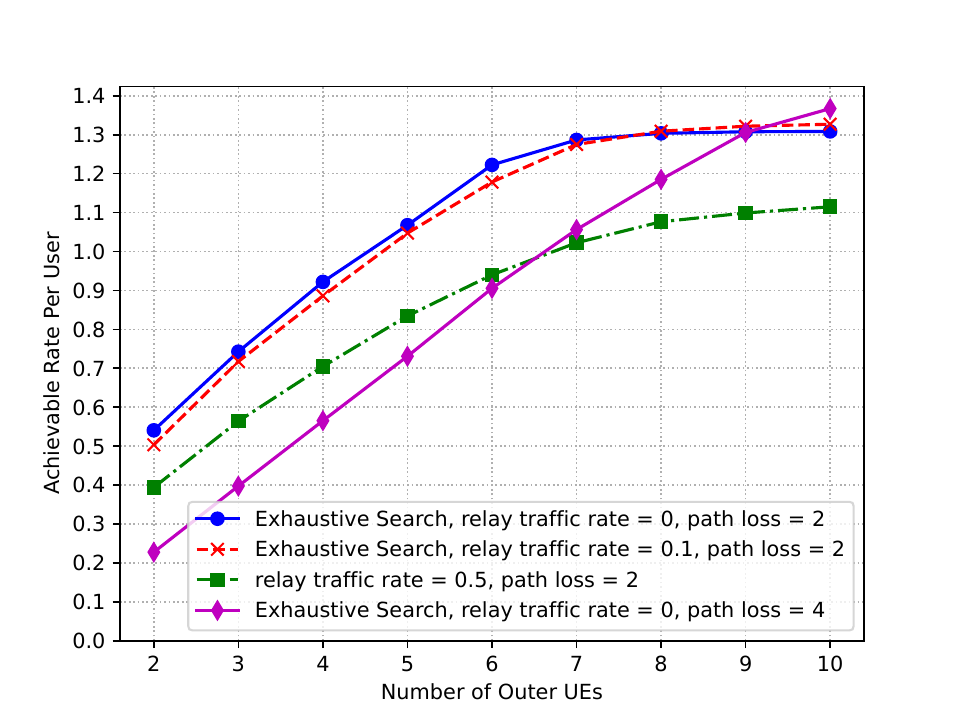}
		\caption{Sum rate achieved by exhaustive search for different properties of relay traffic and path loss. }
		\label{fig:varyrelaytrafficpathloss}
\end{figure}

The resulting achievable sum rate (measured in bit/s/Hz) is shown in Fig.~\ref{fig:sumrate_exhaustive_Random_bc}. The sum rate increases with $n_o$ but only sublinearly. This is expected for interference-limited settings. However, the size of the search space increases exponentially with $n_o$. For the same number of inner and outer UEs, the link schedule is a vector of length $n_o$. Each entry of this vector takes a number from 0 to $n_o$ (0 means UE is not scheduled to transmit). So, the size of search space is $(n_o+1)^{n_o}$. As $n_o$ increases, we cannot practically cover the entire search space and therefore exhaustive search cannot be fully carried out, potentially leading to suboptimal solutions. For example, we have $(n_o+1)^{n_o} < 50000$ (number of searched link schedules) for $n_o \leq 5$ but $(n_o+1)^{n_o} > 50000$ for $n>5$. Therefore, we observe saturation of rate results especially starting with $n_o=6$. In Fig.~\ref{fig:sumrate_exhaustive_Random_bc}, we also show the achievable sum rate for two benchmarks: (i) \emph{random assignments}, where each outer UE is assigned uniformly randomly to any of the available inner UEs and (ii) \emph{best channel assignments}, where each outer UEs is assigned to the available inner UE that has the best channel from the outer UEs based on the received power level. While best channel assignment achieves higher rate than random assignment, both methods remain ineffective as they do not account for interference effects. 

Next, we vary relay traffic rate that is selected randomly from $r_j^{(2)} \in [0,r_{\max}^{(2)} ]$ and vary path loss coefficient $\alpha$. The sum rate (by exhaustive search) is shown in Fig.~\ref{fig:varyrelaytrafficpathloss} for different relay traffic rates. Relay traffic is added to the sidelink communications such that inner (relay) UEs not only carry traffic of outer UEs to the gNodeB but also need to deliver their own traffic to the gNodeB. This effect is reflected in the rate constraint by adding $r_j^{(2)}$, traffic rate from each UE $j \in \mathcal{H}_2$: $r_i^{(1)}+r_j^{(2)} \leq c_{i,b_j^{(2)}}^{(2)}$ for any outer UE $i$  and any inner UE $j=u_i^{(1)}$. The achievable sum rate decreases with the relay traffic, while the optimization accounts for the relay traffic.

The achievable sum rate (by exhaustive search) as a function of $n_o$ is shown in Fig.~\ref{fig:varyrelaytrafficpathloss}, when we increase the path loss coefficient from 2 to 4. In that case, the signal is attenuated more for both intended and interfering signals. Therefore, when $n_o$ is smaller (such that the interference among UEs is smaller), a larger path loss leads to a lower rate as the signal reaching the intended receivers becomes weaker. On the other hand, as $n_o$ increases (such that the interference among UEs increases), a larger path loss leads to less interference power  at the receivers and this effect can compensate the decrease in signal strengths for the intended receivers. As a result, for larger $n_o$, a larger path loss may be beneficial for limiting the interference and improving the achievable sum rate.

\section{Greedy Operation \label{sec:greedy}}
The underlying optimization problem for relay selection in sidelink communications is non-convex (even for single-hop settings \cite{yang2012distributed}) and exhaustive search suffers from the large search space with size increasing with $n_o$. Therefore, in addition to interference, there is the effect of limited search space that shapes the achievable rates. We consider a greedy algorithm that does not suffer from these complexity issues and can scale up with $n_o$, allowing for distributed operation. The problem is to select $u_i^{(1)}$ (assign an outer UE $i$ in to an inner UE) one at a time. Given a set of activated links, adding a link has two expected effects: (i) potential increase in the sum rate since the rate of a new link is added and (ii) potential decrease in the rates of activated links since the new link causes interference to them.

\begin{figure}[ht!]
			\vspace{-0.6cm}
		\centering
		\includegraphics[width=0.8\columnwidth]{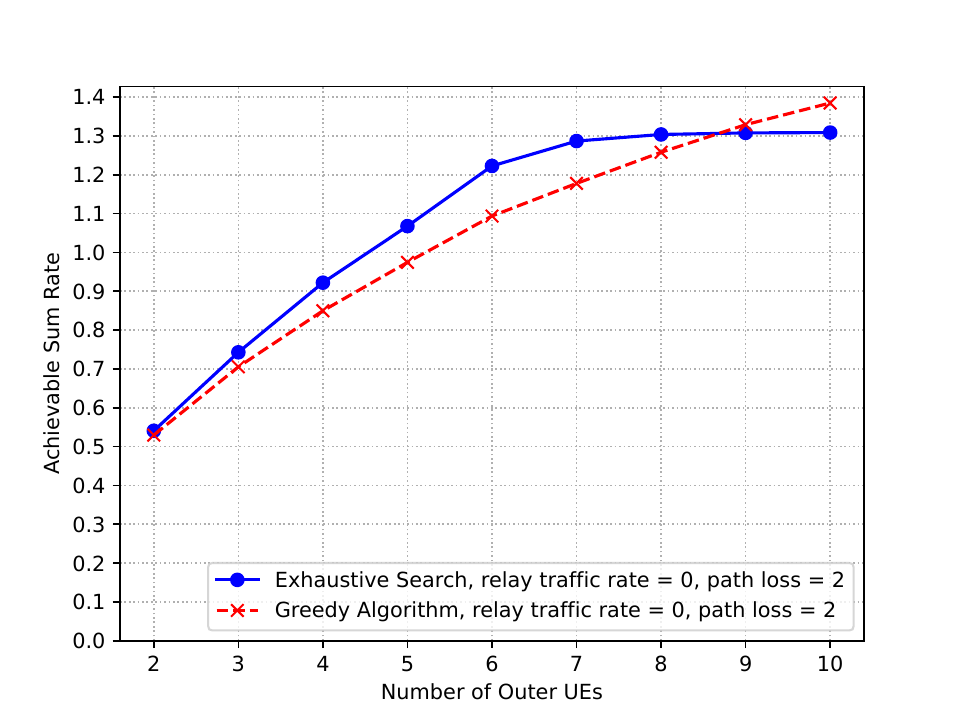}
		\caption{Sum rate achieved by greedy algorithm.}
		\label{fig:greedy}

	\centering
	\includegraphics[width=0.8\columnwidth]{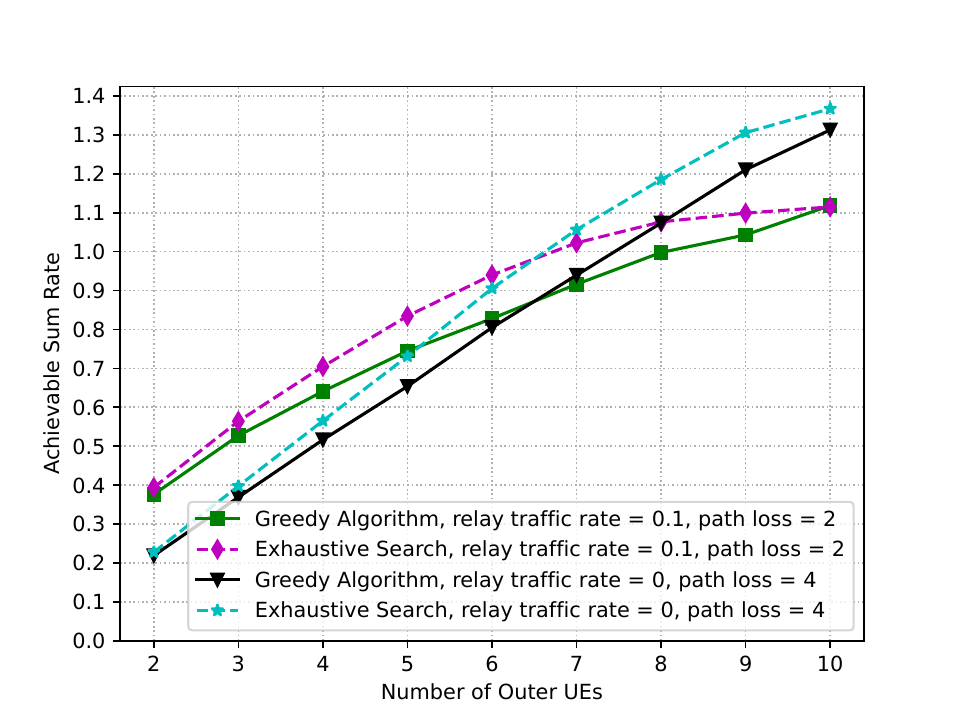}
	\caption{Sum rate achieved by greedy algorithm for different properties of relay traffic and path loss.}
	\label{fig:greedyratepath}

	\centering
	\includegraphics[width=0.8\columnwidth]{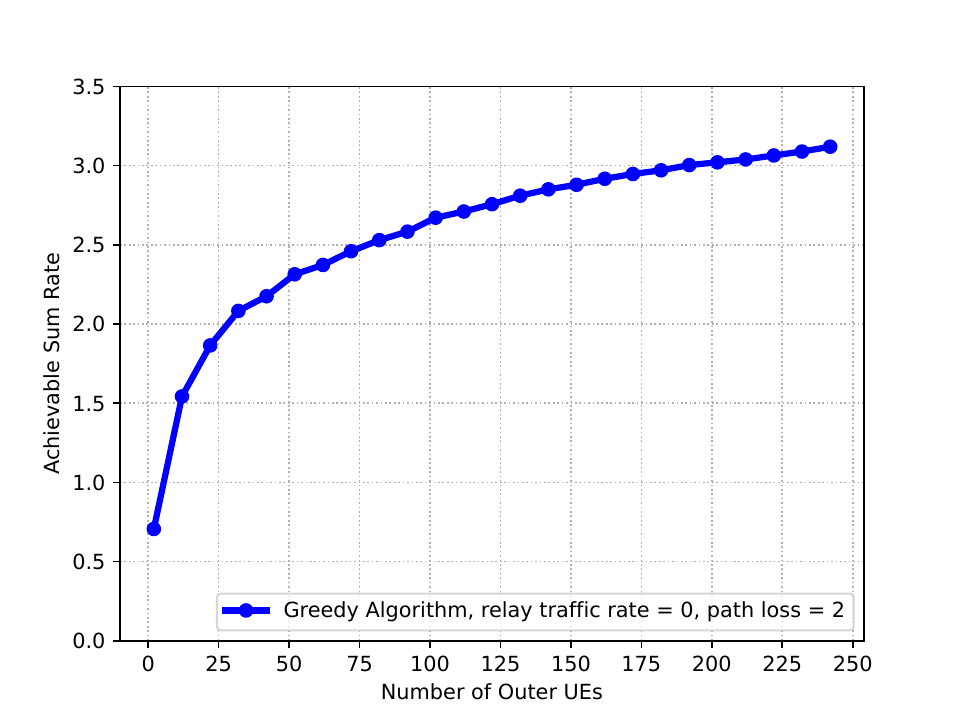}
	\caption{Scaling of sum rate achieved by greedy algorithm with the number of outer UEs.}
	\label{fig:greedyscalable}
	 \vspace{-0.5cm}
\end{figure}

At any time, greedy algorithm adds the best link that maximizes the sum rate given the already activated links. If no such link is found, the remaining (outer) UEs are not activated to transmit. The complexity of greedy algorithm is low and it is computationally easy to implement.  Fig.~\ref{fig:greedy} compares the achievable sum rate (as a function of $n_o$) for exhaustive search and greedy algorithms. The sum rate monotonically increases with $n_o$ under  greedy algorithm. On the other hand, exhaustive search does not scale well with $n_o$ due to limited search space that we can cover, and the increase in sum rate saturates as $n_o$ increases and at that point greedy algorithm achieves higher rates (specifically, when $n_o$ is greater than 8). Fig.~\ref{fig:greedyratepath} varies path loss coefficient and traffic rate of relay UEs, and shows that greedy algorithm can sustain performance under different channel and traffic conditions. In terms of scaling, Fig.~\ref{fig:greedyscalable} shows that the sum rate monotonically increases with $n_o$ and it is computationally feasible to find the link schedule even for a large  number of outer UEs contending access to inner UEs. 

\section{Fair Operation\label{sec:fair}}
\begin{figure}[ht!]
		\vspace{-0.6cm}
	\centering
	\includegraphics[width=0.8\columnwidth]{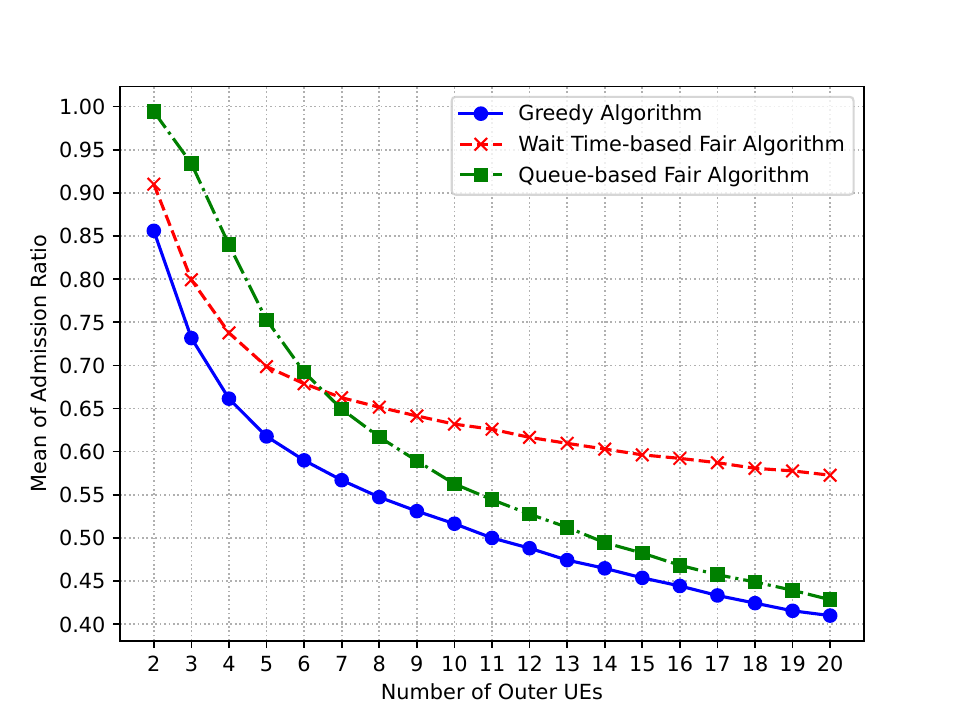}
	\caption{Mean of admission ratio achieved by wait time-based and queue-based fair algorithms.}
	\label{fig:meanadmission}
	\centering
	\includegraphics[width=0.8\columnwidth]{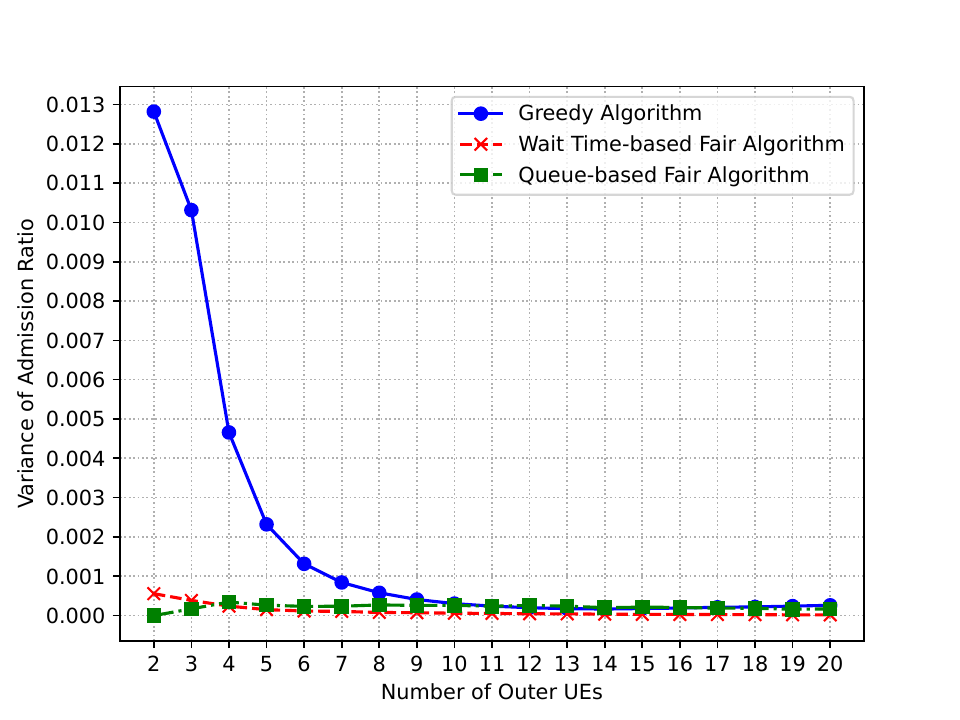}
	\caption{Variance of admission ratio achieved by wait time-based and queue-based fair algorithms.}
	\label{fig:varianceadmission}
	\centering
	\includegraphics[width=0.8\columnwidth]{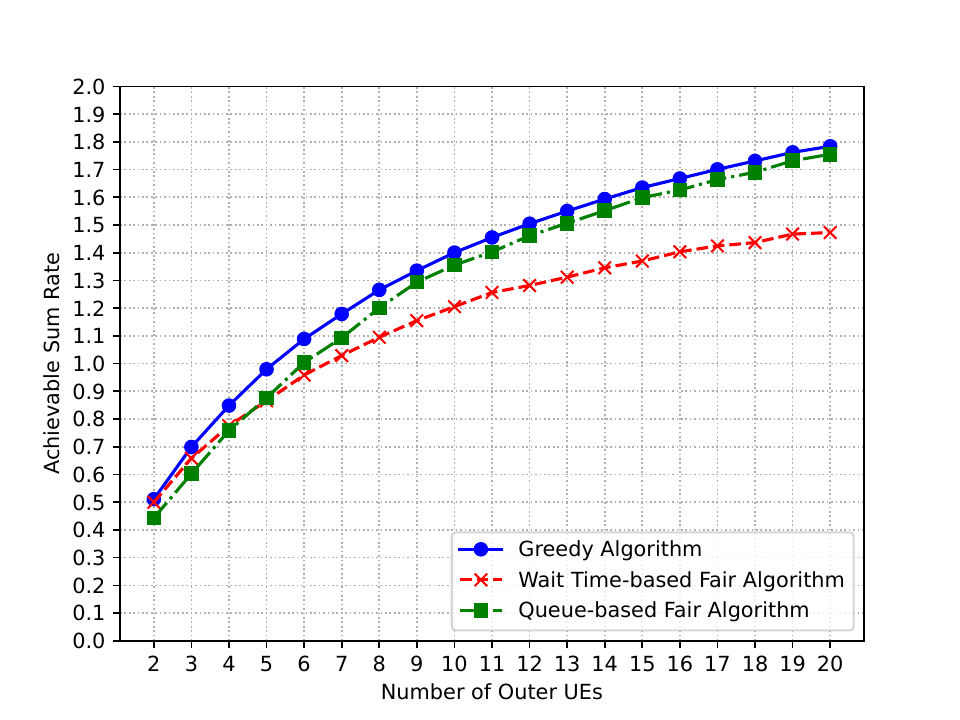}
	\caption{Sum rate achieved by wait time-based and queue-based fair algorithms.}
	\label{fig:sumrateadmission}
	 \vspace{-0.5cm}
\end{figure}

Greedy algorithm may suffer from fairness issue in the sense that given a set of channels, some outer UEs may be scheduled to connect to inner UEs all the time, but some outer UEs may not find this opportunity due to their channel and interference conditions. To understand this fairness issue, one important criterion to measure is the number of outer UEs activated to connect to inner UEs (or the ratio of the number of activated outer UEs to the total number, $n_o$). We measure this admission ratio for outer UEs in terms of its mean and variance across all UEs. Ideally, we should have a high mean for admission ratio so a large number of outer UEs can be served and we should have a low variance for admission ratio across all outer UEs  so that we have a small difference across outer UEs in terms of the ratio of time they can be served, leading to a fair operation. In Figs.~\ref{fig:meanadmission} and \ref{fig:varianceadmission}, we show the mean and variance of admission ratio of the number of outer UEs connected to inner UEs under greedy algorithm.

We observe that as $n_o$ increases, it becomes more difficult to serve them due to the increasing interference effects that are expected to reduce the achievable sum rate. Therefore, the mean of admission ratio decreases with $n_o$, i.e., it becomes more difficult to serve the increasing number of UE connection requests. On the other hand, the variance of admission ratio across outer UEs also decreases with $n_o$, i.e., greedy algorithm provides fairer service when there are more UEs to serve. A better channel is not the sole criterion in link selection of greedy algorithm. The best link is iteratively added (one by one) by also accounting for interference effects (from and to established links). The optimization problem formulation allows adding weights to rates to support priorities in addition to channel conditions. Specifically, a weight can reflect the waiting time for each outer UE since its last activation to transmit to an inner UE. If some outer UEs with good channel or interference properties are selected, this would build waiting time of admission for other outer UEs and then their weights increase over time, and eventually they will get selected by greedy algorithm. To reflect this wait time feature, we assume a slotted system and each UE $i$ updates its weight $w_i$ at each time slot as follows: 
\begin{enumerate}
\item $w_i \leftarrow w_i + 1 $, if UE $i$ is not activated to transmit in a given time slot, or
\item $w_i \leftarrow \max(w_i - 1, 1) $ if UE $i$ is activated to transmit in a given time slot.
\end{enumerate}

We can also consider the case when service requests are arriving stochastically as packets and queued over time. The arrival rate for queue of UE $i$ is $\lambda_i$ and the service rate is $r_i$ achieved at the given time slot. To reflect queueing dynamics, each UE $i$ updates its weight $w_i$ at each time slot as follows: 
\begin{enumerate}
	\item $w_i \leftarrow w_i + \lambda_i $, if UE $i$ is not activated to transmit in a given time slot, or
	\item $w_i \leftarrow \max(w_i + \lambda_i - r_i, 0)  $ if UE $i$ is activated to transmit in a given time slot.
	\end{enumerate}
	
An important fairness criterion is the admission ratio, namely the ratio of outer UEs allocated for transmission over all outer UEs. It is desirable to have both a large mean of admission ratio (so that more UEs can be served on the average, where the average is taken over time and outer UEs) and a small variance of admission ratio across outer UEs (so that outer UEs are fairly served). For numerical results, we assume that the operation takes 500 slots and is repeated for 1000 times for averaging the simulation instances, and consider Bernoulli process for packet arrivals. Figs.~\ref{fig:meanadmission}, \ref{fig:varianceadmission}, and \ref{fig:sumrateadmission} show the mean of admission ratio, the variance of admission ratio, and the achievable sum rate, respectively. Fair algorithms achieve higher mean of admission ratio and lower variance of admission ratio compared to greedy algorithm. Mean of admission ratio decreases with $n_o$ since the increase in interference limits the achievable sum rate. Variance of admission ratio remains high for greedy algorithm and drops to the level of fair algorithms only for large $n_o$. Fig.~\ref{fig:sumrateadmission} shows that greedy algorithm can sustain higher achievable sum rate as it is designed to do, whereas fair algorithms, especially queue-based algorithm, can reach close to these rate levels.

\begin{figure}[t!]
		\vspace{-0.5cm}
	\centering
	\includegraphics[width=0.8\columnwidth]{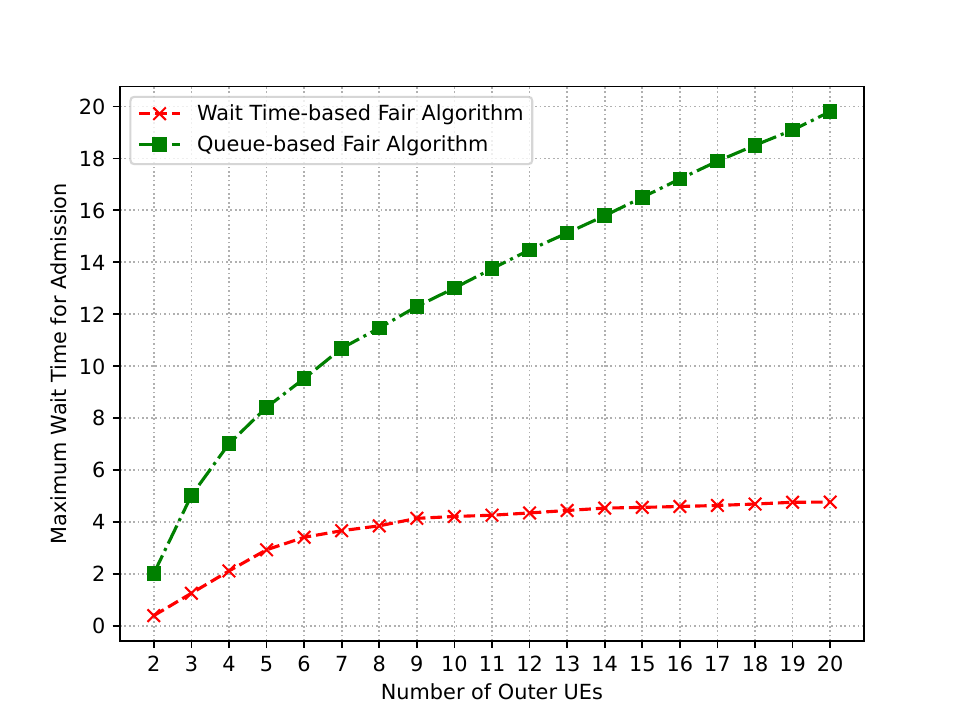}
	\caption{Maximum wait time incurred by wait time-based and queue-based fair algorithms.}
	\label{fig:waittimeadmission}	
		\centering
	\includegraphics[width=0.8\columnwidth]{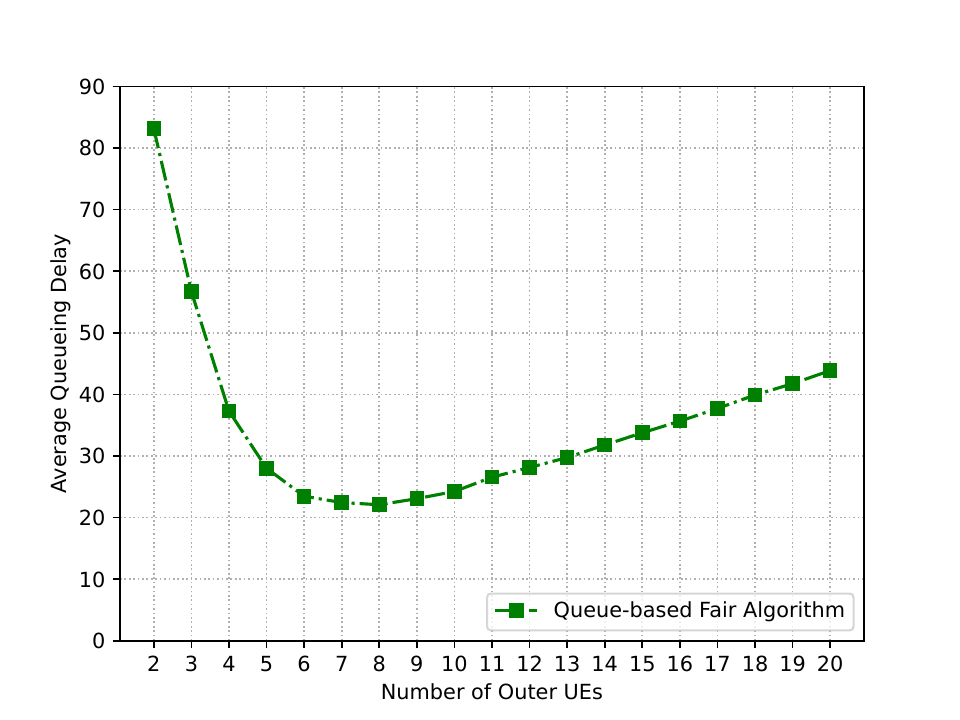}
	\caption{Average queueing delay incurred by queue-based fair algorithms.}
	\label{fig:delayadmission}
\end{figure}

Next, we consider the wait times from one activation to another one and evaluate the maximum value incurred by all UEs over time. Greedy algorithm may not at all activate some of UEs, for which the wait time cannot be assessed.  Fig.~\ref{fig:waittimeadmission} shows the maximum wait time for fair algorithms. The maximum wait time remains low for the wait time-based algorithm, whereas it quickly increases for the queue-based algorithm as $n_o$ increases. 

Finally, we evaluate the queueing delay for the queue-based fair algorithm. We assume that the number of arrivals is uniformly randomly distributed such that the mean of total arrival rate across outer UEs is $0.5$. Fig.~\ref{fig:delayadmission} shows how the average delay varies with $n_o$. When there are only few UEs, the arrival rate per UE is high compared to service rate so the queueing delay is high. When $n_o$ increases, the average delay decreases first since the service rate increases more relative to arrival rate, reaches its lowest value (when $n_o$ is 8), and then the interference starts limiting the increase in service rate and more UEs lack activation over time such that the average delay starts increasing with higher $n_o$.

This approach offers scheduling over multiple hops including UE-UE and UE-gNodeB associations by incorporating two-hop routing and activation of UE transmissions. By considering differential backlogs of multiple traffic flows (for different source UEs) as weights, this approach can be potentially extended via backpressure routing \cite{tassiulas1992stability} to mobile ad hoc network (MANET) routing over more than two hops. 

\section{Distributed Operation \label{sec:distributed}}
For practical implementation, we consider distributed operation of relay selection in sidelink communications. Channel State Information Reference Signal (CSI-RS) is used for channel state estimation/sounding and reporting between a transmitter and a receiver UE. To establish outer-inner UE connection, outer UEs that want to reach a gNodeB transmit ``\textit{I am here}" discovery messages at pre-defined discovery intervals. Inner UEs directly connected to the gNodeB monitor the spectrum and process these messages once received. Specifically, the following procedure is pursued: 

\begin{enumerate}
	\item Outer UEs transmit ``\textit{I am here}" messages.
	\item Inner UEs send ACK ``\textit{I hear you}" messages for all received ``\textit{I am here}" messages. If multiple UEs transmit at the same time, their messages may collide (no ACK) and undergo through a CSMA-style backoff mechanism.
	\item Inner UEs measure the CSI from all received messages and deliver it to the gNodeB.
	\item gNodeB runs greedy algorithm at any given time and determines with UE links to add.
	\item gNodeB broadcasts these assignments to all inner UEs.
	\item Inner UEs send a second ACK message to the outer UEs to be activated. 
\end{enumerate}

If there are $n_o$ outer UEs and $n_o$ inner UEs), the number of messages exchanged for distributed algorithm is bounded as follows: $n_o$ for step 1, $n_o$ for step 2 (if no collision resolution is needed), $n_o$ for step 3, 1 for step 5, and $n_o$ for step 6. Overall, $4 n_o+1$ messages need to be exchanged (excluding messages needed to resolve potential collisions in Step 2).

\section{Conclusion} \label{sec:conclusion}
In this paper, we studied relay selection and UE admission in sidelink communications for emerging 5G/6G networks, focusing on enhancing connectivity and managing network resources more effectively. Through the development of greedy and fairness-oriented algorithms, we addressed both efficiency in network performance and equity in UE access to the network. Results show promising  enhancements in network connectivity and user inclusivity in emerging 5G/6G networks.

\bibliographystyle{IEEEtran}
\bibliography{references}

\end{document}